\documentclass[10pt,journal]{IEEEtran}

\usepackage{url}
\usepackage{graphicx}
\usepackage{array}
\usepackage{color}
\usepackage{times}
\usepackage{comment}
\usepackage{nomencl}
\usepackage{amsmath,amsfonts,amssymb,mathrsfs}
\usepackage{subfigure}
\usepackage{algcompatible}
\usepackage{algorithm}
\usepackage[table]{xcolor}
\usepackage{multirow}
\usepackage{booktabs}

\begin{document}

\title{Data Integrity Threats and Countermeasures in Railway Spot Transmission Systems}

\author{Hoon~Wei~Lim, 
        William~G.~Temple, 
        Bao~Anh~N.~Tran, 
        Binbin~Chen, 
        Zbigniew~Kalbarczyk 
        and Jianying~Zhou 
\thanks{H.W. Lim is with Singtel. The work was done while the author was working at the Institute for Infocomm Research (I2R), Singapore.}
\thanks{W.G. Temple and B. Chen are with the Advanced Digital Sciences Center (ADSC), Illinois at Singapore, Singapore.}
\thanks{B.A.N. Tran is with Wargaming, Sydney, Australia. The work was done while the author was with ADSC.}
\thanks{Z. Kalbarczyk is with the University of Illinois, Urbana-Champaign.}
\thanks{J. Zhou is with the Singapore University of Technology and Design (SUTD).}}

\maketitle

\begin{abstract}
Modern trains rely on balises (communication beacons) located on the track to provide location information as they traverse a rail network. Balises, such as those conforming to the Eurobalise standard, were not designed with security in mind and are thus vulnerable to cyber attacks targeting data availability, integrity, or authenticity. In this work, we discuss data integrity threats to balise transmission modules and use high-fidelity simulation to study the risks posed by data integrity attacks. To mitigate such risk, we propose a practical two-layer solution: at the device level, we design a lightweight and low-cost cryptographic solution to protect the integrity of the location information; at the system layer, we devise a secure hybrid train speed controller to mitigate the impact under various attacks. Our simulation results demonstrate the effectiveness of our proposed solutions.
\end{abstract}
\section{Introduction}

In railway systems, communications-based train control (CBTC)~\cite{IEEE1474,PE09} and positive train control (PTC)~\cite{PF12} systems are widely adopted and deployed in the Europe and the U.S., respectively.
This is to ensure a safe distance between two (successive) moving trains. 
In particular, knowledge of the accurate location of a front train is critical for the train that follows behind, so an appropriate speed limit and braking distance can be estimated and imposed on the back train.
This enables shortening of headways between successive trains.

\subsection{Communications-based Train Control (CBTC)}

The principal intent of a train control system is to prevent collisions when trains are traveling on the same track, either in the same or opposite direction.
CBTC was designed to minimize headway, improve flexibility and utilization through the concept of {\em moving blocks}---giving the following train a movement authority up to the exact rear-end location of the lead train~\cite{PE09}. 
This, in turn, can be realized through a combination of: (i) accurate train location tracking based on {\em balises}\footnote{A balise, which is a passive RIFD tag, is originally a French word to distinguish a beacon for railway than other beacons.} (also known as beacons) mounted between rails; (ii) continuous track-to-train and train-to-track communication of control and status information; and (iii) trackside and train-borne vital processors to process the train status and control data and provide continuous automatic train protection (ATP)~\cite{IEEE1474}.
Typically, an ATP unit determines the permissible safe speed profile of a train on which it is installed.
An automatic train operation (ATO) unit then controls the train under the supervision of the ATP.
The latter controls the movement of the train by making use of geographical data defining physical characteristics of a railway track, and location data associated with the routes to be taken along the track by the train~\cite{Newman95}.

Generally, CBTC is developed specifically for metro (urban) or high-capacity railway systems.
It is not a “standard” product, but mostly proprietary.
In contrast, the European Rail Traffic Management System (ERTMS) and European Train Control System (ETCS)~\cite{ERTMS} are developed for main lines (inter-city or suburban) to address railway interoperability requirements mandated by the European Railway Agency.
Nevertheless, commercial CBTC solutions by leading railway signaling vendors, such as Alstom~\cite{Urbalis}, Siemens~\cite{Trainguard} and Thales~\cite{SelTrac}, are typically compliant with the ERTMS/ETCS standards. 

\subsection{Location Tracking}

An integral part of CBTC (and ETCS) is a spot transmission system that provides train location information with high-degree of precision.
This is realized through transmission of location data from a balise to a passing train.
Balises need to be powered up by a passing train in response to radio frequency energy broadcast by a balise transmission module (BTM) mounted under the train.
Such data transmission is {\em intermittent} on the order of fractions of a second.
Moreover, no handshake is required, so that the BTM is able to receive as much data as possible when a train passes a balise even at speed as high as 500 km/h.

Modern balise technology, introduced in early 1990s, was pioneered by Alstom (French KVB) and Siemens (German ZUB).
The development of the Eurobalise standard was largely influenced by them and some other vendors such as Ansaldo, Bombardier, and Thales.
Hence, the design of telegrams and their usage for Eurobalise have been continuously refined overtime and well-tested to meet high-level of safety requirements.
However, while the Eurobalise technology has been designed to be safe, security concerns, such as false data injection and replay attacks, have never been seriously considered.
This is perhaps because the art and science of cybersecurity was previously not yet well-understood when Eurobalise was initially designed. 

\subsection{Contribution}

In this paper, we study the Eurobalise spot transmission system~\cite{Subset36} that is adopted by ERTMS/ETCS and commercial CBTC solutions.
Particularly, we examine the encoding and decoding schemes designed for the Eurobalise standard and provide insights on potential security issues related to intermittent wireless communications.
(Note that we focus on Eurobalise because of its internationally high adoption rate and that its design details are publicly available.)
We identify open technical challenges that need to be addressed to improve the security of the Eurobalise spot transmission system.

To address those challenges, 
we propose a {\em lightweight} and {\em low-cost} cryptographic solution to protect the location integrity of a train.
In particular, we embed authentication code into specific parameters used in Eurobalise telegrams such that any intentional or unintentional changes to a telegram's user data would be detected. Our approach is lightweight in the sense that the additional computational overhead introduced by our technique is negligible compared to the allowance permitted in the current standard for CBTC systems~\cite{IEEE1474}.
Furthermore, our approach is low-cost because no additional hardware or devices are required compared to the existing system. 

Finally, to provide an additional layer of security assurance for train operation, we leverage prior work on control-theoretic 
countermeasures for balise data attacks~\cite{Temple2017} and integrate the cryptographic countermeasure into a secure hybrid train speed controller. Integrating those techniques preserves the advantages of the telegram authentication scheme while providing an operationally-viable response strategy when data integrity attacks are detected. We demonstrate the performance of this hybrid countermeasure through simulation using the case study of train automatic stop control. 


\section{Balise Background} 
\label{sec:background}

In this section, we give an overview of the design of the Eurobalise spot transmission system as specified in the UNISIG Standard SUBSET-036~\cite{Subset36}. We also introduce balise applications in train control. 

\subsection{System Design}
\label{sec:system-design}

Spot transmission in our context refers to a phenomenon when a transmission path exists between the trackside and the on-board BTM at discrete locations.
The location data (pre-programmed and stored on a balise) is emitted and picked up by a passing train only as its antenna unit passes over the corresponding balise.
The length of track on which the data is passed and received is limited to approximately one meter per balise.

\begin{figure}[h!]\centering
	\includegraphics[width=0.28\textwidth]{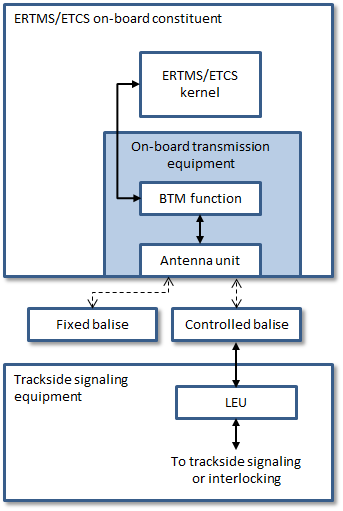}
	\caption{Key components of the Eurobalise transmission system~\cite{Subset36}.}
	\label{fig:BTM}
\end{figure}

Figure~\ref{fig:BTM} shows the architectural design of the Eurobalise spot transmission system.
It comprises three major components: on-board transmission equipment, balise, and trackside signaling equipment.
Generally there are two types of balises: {\em fixed} and {\em controlled}. However, from a vendor point of view the devices are largely the same and this distinction is a matter of configuration (e.g., the two balise models in~\cite{Trainguard} are available as either fixed or controlled).

A fixed balise is programmed to transmit the same data in the form of a telegram to every train.
Fixed balises typically transmit information such as the location of the balise, geometry of the line (e.g., curves and gradients), and any speed restriction(s) in that particular area. 
On the other hand, a controlled balise is used when the spot transmission system is required to be overlaid onto a traditional (national or mainline) signalling system.
The controllable balise is typically connected to a lineside electronics unit (LEU), which transmits dynamic data to the train.
The LEU then integrates with the conventional signaling system either by connecting to the lineside railway signal or to the signalling control tower.\footnote{The LEU obtains variable data from the interlocking control system signal on aspects, train location, temporary speed restrictions, and etc.}

The main role of the on-board transmission equipment is to provide support for balise localization. 
It uses an amplitude modulation on the 27.095 MHz frequency to power the passive balises, typically known as {\em tele-powering}.
In addition to generation of tele-powering signal, the on-board transmission equipment provides a range of safety functionalities, including detection of bit errors, immunity to environmental noise, and physical {\em cross-talk}\footnote{The term ``cross-talk'' refers to a situation when a telegram is read from a balise that should not be read, e.g., balise on another track.} protection.
Further details on the aforementioned functionalities can be found in~\cite{Subset36}. 

\subsection{Applications in Train Control}
Balises, whether fixed or controlled, are a vital component of automatic train operation and automatic train protection systems. Balises may be deployed individually (e.g., to provide a reference position) or in groups of up to eight to communicate a larger quantity of information to a passing train~\cite{Subset36}. In main line systems, balises are often deployed in pairs along the track at regular intervals (on the order of 1km apart)
to support train localization and help the train infer its direction of travel. 

In automated train systems, such as a metro or an airport people mover, balises serve another important purpose: helping the trains stop correctly at the station. More specifically, balises placed near the station help the train determine when to start braking and when it has reached the desired stop point. In systems with platform screen doors, which prevent passengers from accidentally falling into the track area, the stopping requirements are stringent (on the order of tenths of a meter) 
since the train doors and platform doors must align. 

\section{Balise Security Challenges and Threats} 
\label{sec:challenges-main}

Traditionally, railway systems have been designed with safety in mind. 
Hence, current railway systems are generally thought to provide a sufficiently high-level of safety for their passengers, and it may be difficult, if not impossible, to circumvent these safety mechanisms by exploiting vulnerabilities in the cyber aspect of railway systems.
However, with the advancement and availability of wireless communication technology, it is unclear what could be the security impact to a train spot transmission system.


Bloomfield et.~al.~\cite{BBG+12} conducted a security assessment on ERTMS/ETCS~\cite{ERTMS}, which includes the Eurobalise system.
Here is an excerpt from their findings:
\begin{quote} \small
	``Messages from balises are protected against accidental transmission errors and interference from outside the immediate area of the track. 
	However, the interface does not address the possibility that an attacker might have subverted existing balises, or placed a new balise on the track at a strategic location. 
\end{quote}

An independent study by Bezzateev, Voloshina, and Sankin~\cite{BVS13} shows that there is no concerted effort in developing not only safe, but also secure railway systems, in the current ERTMS/ETCS standards.

We note that balises can now be activated and deactivated by a wireless device, such as Alstom's Balise Encoder Programming and Test (BEPT) tool~\cite{Atlas} and Siemens's TPG Eurobalise V2~\cite{TPG-Eurobalise}.
This feature, partly motivated by safety considerations, allows railway maintenance staff to work remotely away from the track to carry out programming of telegrams on both fixed and controlled balises. 
However, conversely, this also opens up the possibility of unauthorized re-programming of telegrams by a malicious insider or an attacker in possession of such a tool with the intent of causing a major train disruption.


\subsection{Threat Model}
\label{sec:threatmodel}
In this paper, we consider data integrity and authenticity threats arising from balise reprogramming devices, as shown in Figure~\ref{fig:balise-programming}.
Specifically, we assume that an adversary is able to obtain or replicate a balise reprogramming device, gain physical access to balises on the track, and implement one of the following attacks:
\begin{figure}[t!]\centering
	\includegraphics[width=0.49\textwidth]{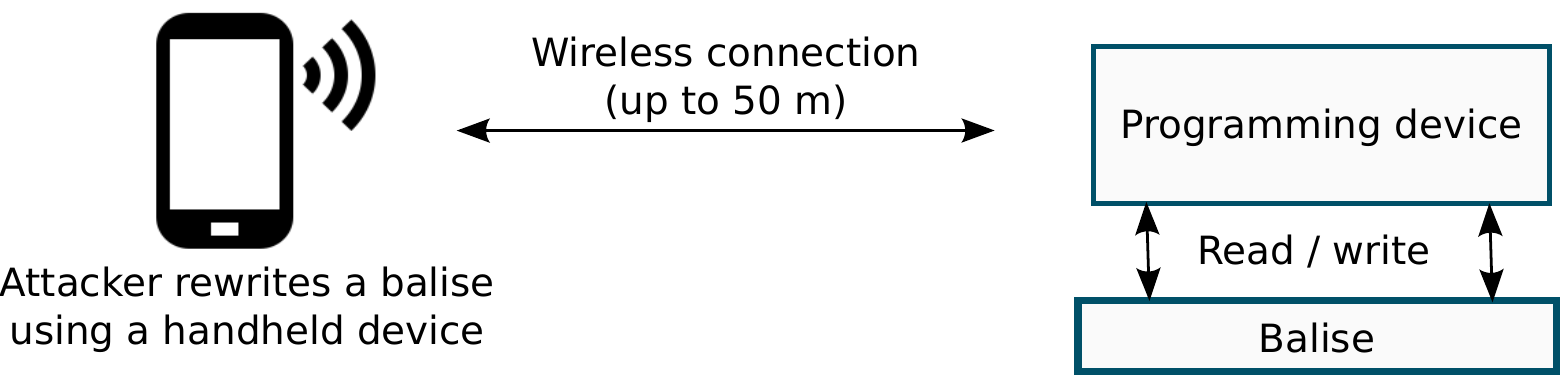}
	\caption{A reprogramming attack using a wireless balise programming tool.}
	\label{fig:balise-programming}
\end{figure}
\begin{itemize}
\item \textbf{Tampering Attack:} modifying or re-writing the telegram on a balise to inject false data.
\item \textbf{Cloning Attack:} a special case of the Tampering Attack where the attacker copies a valid telegram from one balise onto another. 
\end{itemize}
In addition, we assume that the attacker is knowledgeable about railway systems and operating principles. Although we focus on capability rather than intent, one could imagine the above attacks being initiated by a disgruntled employee or contractor, or a determined outsider (e.g., industrial espionage). Since the outcome of a successful balise integrity attack is likely to be a service disruption rather than a safety incident, the negative outcome for a rail transit operator is likely to be reputation damage (e.g., negative press, unhappy commuters) or monetary loss (e.g., fines based on quality of service metrics). 

Note that we do not consider attacks involving the redeployment of existing balises (e.g., removing a balise from one location and reinstalling it elsewhere). However, the attacker may compromise multiple balises using the tampering and/or cloning attacks. After countermeasures are introduced (see Section~\ref{sec:device-level}), the attacker does not have access to the cryptographic key material used to protect data integrity. 
\subsection{Case Study: Train Automatic Stop Control}
\label{sec:usecase}
While the threat model presented above is applicable to any train system using a Eurobalise transmission system, we select a specific case study to provide an illustration of the impact of attacks and the efficacy of our countermeasures.
We select a train automatic stop control case study that has been used in prior work to examine availability attacks targeting balise transmission~\cite{Temple2017}. The model is based on real systems in use today, as discussed in~\cite{Chen2013balise}. 
\begin{figure}[h!]\centering
	\includegraphics[width=0.9\linewidth]{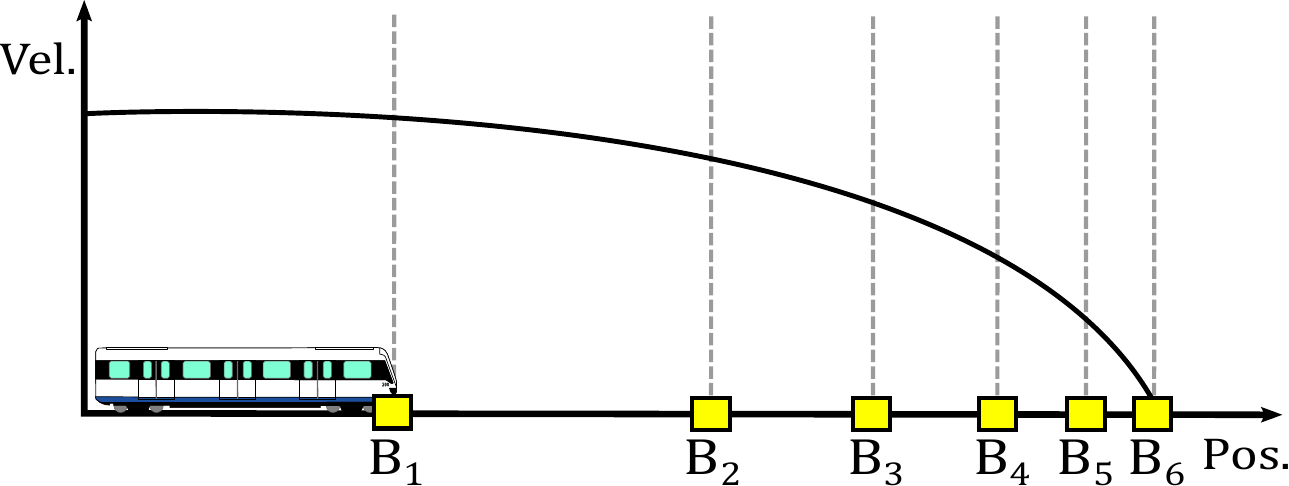} 
	\caption{Illustration of a train stop with position input from fixed balises.}
	\label{fig:tasc}
\end{figure}
\begin{figure*}[!t]
	\centering 
	\includegraphics[width=0.8\linewidth]{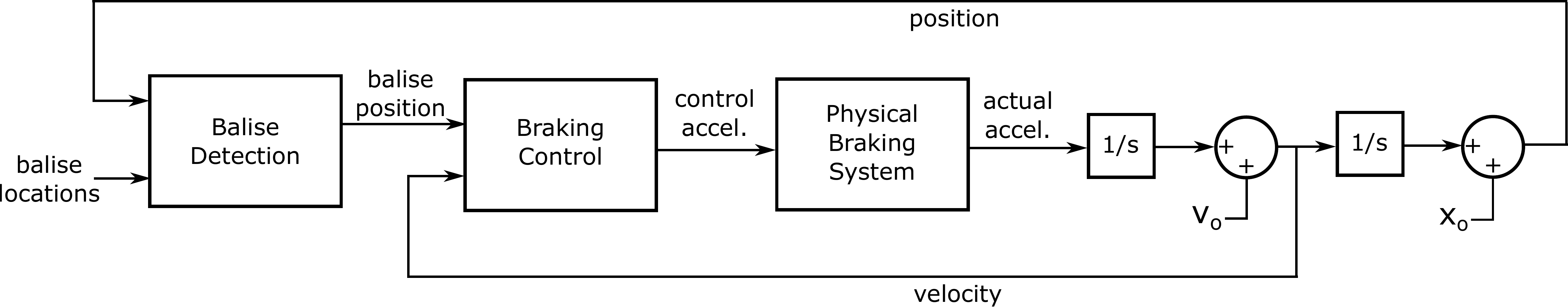}
	\caption{Block diagram of the stop control model with no countermeasures against balise attacks.}
	\label{fig:blockdiagram}
	\vspace{-3mm}
\end{figure*}

In this case study a driverless train approaching a station controls its braking action using balise position references to stop at a desired position (see Figure~\ref{fig:tasc}). There are a series of $m$ balises at a station, $B = \{B_1, B_2, \ldots, B_m\}$. Note that the number $m$ and spacing of balises may vary between train systems depending on a vendor's technology and the application requirements. A stop is considered successful if the train is within $\pm \gamma$ from the desired stop point $B_m$. 
Each balise $B_i$ is represented as a tuple $B_i = <loc_i, loc^\prime_i>$ where $loc_i$ is the physical location of the balise and $loc^\prime_i$ is its reported location in the telegram. Positions are defined relative to the stopping point: in other words $loc_m=0$, and $loc_1 < loc_2 < \ldots < loc_m$. 


A number of train stop control algorithms have been proposed and/or implemented in operational train systems. These include conventional proportional-integral-derivative (PID) controllers, online learning algorithms~\cite{Chen2013balise}, and the use of pre-calculated velocity profiles (run curves) or upper and lower velocity bounding curves (see~\cite{DiCairano16} and references therein). 
In this work we consider a learning-based braking controller. As the train passes over balise $B_i$ it will adjust its deceleration rate $\alpha$ as it tries to achieve a full stop at $B_m$. Adjustment is needed due to the various disturbances (e.g., actuator delay, friction, wheel slip) a train may encounter as it slows. 
Further detail about the braking control model is given in~\cite{Temple2017}. To provide context for the results presented in Section~\ref{sec:evaluation}, we summarize key elements of the model below. \newline


\noindent\textbf{Control Algorithm} For this case study we employ the heuristic online learning algorithm (HOA) in~\cite{Chen2013balise}, 
which has been successfully applied in an operational subway system over several years. 
%
In this model there are three key parameters:
the expected deceleration without disturbances ($\alpha_i^e$), the controller deceleration ($\alpha_i^c$), and the estimation of the actual deceleration ($\alpha_i^r$) realized by the train. 
The expected deceleration as a train passes a balise is obtained from the equations of motion: 
\begin{equation}
\alpha_i^e = \frac{v_i^2}{2loc^\prime_i}.
\end{equation}
Note that this takes a negative value since $loc^\prime_i$ is negative with respect to the zero point.  
The controller deceleration is obtained by adjusting $\alpha_i^e$ in response to the observed train deceleration. Specifically,
\begin{equation}
\alpha_{i+1}^c = \alpha_{i+1}^e - \eta_i (\alpha_i^r-\alpha_i^c), 
\end{equation}
where
\begin{equation}
\eta_{i+1}= 
\begin{cases}
0.95 \times \eta_i & \text{if } |\alpha_i^r-\alpha_i^c| > 0.05 \\
1.05 \times \eta_i & \text{if } |\alpha_i^r-\alpha_i^c| \leq 0.05
\end{cases}
\end{equation}

$\alpha_i^r$ is calculated using $v_i$, $v_{i+1}$, and the distance between successive balises 
\begin{equation}
D_i = |loc^\prime_i| - |loc^\prime_{i+1}|.
\end{equation}
The acceleration is given by:
\begin{equation}
\label{eq:alphai}
\alpha_i^r=(v_{i+1}^2-v_i^2) /2D_i.
\end{equation}
The actual train deceleration, $\alpha_i^{train}$, is influenced by various disturbances and system implementation features. We use a train braking model which was empirically determined based on data from an operating subway system~\cite{Chen2013balise}. 
The actual deceleration of the train is given by the transfer function:
\begin{equation}
\alpha^{train} = \frac{\alpha_0}{T_ps+1}e^{-T_ds}
\end{equation}
where $T_d$ is the system's time delay and $T_p$ is the time constant.
The overall control model, which is implemented in Matlab/Simulink, is illustrated in Figure~\ref{fig:blockdiagram}. In Section~\ref{sec:evaluation} we simulate this stop control process with and without attacks on balise data to evaluate the countermeasures presented in the following sections. 

\section{Related Work} 
\label{sec:localization-state-of-the-art}

In this section we give an overview of related prior work on secure localization, cryptographic checksums, and commercial train localization solutions to provide context for the countermeasures introduced in Sections~\ref{sec:device-level} and~\ref{sec:system-level}.

\subsubsection{Secure Localization of Moving Objects}

It is well-known that the coordinates of a moving object, e.g., train, car, or airplane, can be determined with a relatively high precision with the aid of at least three verifiers (receivers).
The latter are typically satellites that are capable of individually calculating the position of the moving object.
This is generally known as location verification through {\em multilateration}, an approach taken by~\cite{Franckart04,SLM15,SLS15}, for example.
However, such a system was typically not designed to prevent unauthorized or intentional modification to the location data sent to the verifiers.
Hackers and academic researchers~\cite{Kunkel10,CF12,SLM13} have demonstrated the feasibility of various message injection and manipulation attacks against the multilateration system used for next generation air traffic control, i.e., Automatic Dependent Surveillance - Broadcast (ADS-B).
Furthermore, deployment of navigation satellites or the like could be expensive, and multilateration is highly susceptible to measurement errors due to noisy environments.

\subsubsection{Cryptographic Checksums}

Checksum algorithms can range from simple parity bits, fingerprints, hash functions, to more advanced error-detecting codes, such as cyclic redundancy codes (CRCs)~\cite{Peterson61}, and error-correcting codes, such as Reed-Solomon codes (RSCs)~\cite{RS60}.
An error-correcting code (ECC) not only detects common errors, but also recovers the original data in some cases when using appropriate parameter choices.
However, while checksums are often used to verify data integrity, they should not be relied upon for verifying data authenticity. 

In cryptography, HMAC~\cite{BCK96} and CBC-MAC~\cite{BKR94} are widely used MACs in real world applications.
Nevertheless, MACs are not designed to detect and correct errors, such as random or burst errors, introduced during transmission as with what ECCs do.
Consequently, various techniques for cryptographically secure checksum algorithms, which concurrently detect errors and ensure data authenticity/integrity, have been proposed~\cite{Rabin81,Krawczyk94,Shoup96}.
A notable result is that a MAC may be constructed  as the composition of a universal hash function (UHF) with a pseudorandom function (PRF).\footnote{A UHF is an algebraic function~\cite{CW79} that compresses a message or data block into a compact digest based on a key.
However, it is not a cryptographically secure primitive.}

It turns out that ECCs such as RSCs can be suitable candidates for UHFs.
As shown by Bowers, Juels, and Oprea~\cite{BJO09}, it is possible to embed MACs in the parity blocks of ECCs, such that a block is simultaneously both a MAC and a parity block.
They called this integrity-protected error-correcting code (IP-ECC), a key enabler of their proposal of a distributed cryptographic solution for high-availability and integrity-protected cloud storage system.
Independently, error-tolerant MACs based on Bose-Chaudhuri-Hocquenghem (BCH)~\cite{BR60} and Reed-Solomon error-correcting codes have been investigated in~\cite{LGV02,GAD06}.
More recently, Dubrova et.\ al.~\cite{DNS15,DNSL15} proposed techniques for constructing cryptographically secure CRCs based on different variants of generator polynomials.

Despite the above advancement in secure cryptographic checksum algorithms, these techniques cannot be directly applied to address our security concerns.
This is due to various unique physical and design constraints that exist in a train spot transmission system (to be further discussed in Section~\ref{sec:challenges}).

\subsubsection{Commercial Patented Technologies}

We have examined relevant granted patents owned by various railway signaling vendors, including Alstom~\cite{Franckart04}, Ansaldo STS~\cite{HL12}, General Electric~\cite{Fries14}, Thales~\cite{Kanner13,MK15}, and Wabtec~\cite{Kernwein12}. 

Our findings are that the inventions disclosed in these patents aim to improve on the accuracy and reliability of a train location tracking system.
Generally, these inventions make use of one or a combination of the following methods: 
\begin{itemize}
	\item Multilateration through sensors or navigation satellites;
	\item Track map;
	\item Challenge-response protocol;
	\item Groups of transponders.
\end{itemize}
However, there exist trade-offs between deployment cost, maintainability, accuracy/reliability.
In comparison with our approach, the above methods seem to be either much more costly, or more challenging to deploy and maintain. 
For example, Alstom has a method for secure determination of an object location, particularly a vehicle moving along a known course using multilateration based on at least 4 navigation satellites (including radio and sensor), and relying on a secure (integrity-protected) on-board mapping database for cross-checking purposes. 
The latter may grow quickly if the database also stores data such as track geometry and feature geo-locations, and thus, may not be easy to maintain and high-rate bandwidth is required to upload/update the database to an on-board storage equipment.

\section{Integrity Protection: Device Level}
\label{sec:device-level}

In this section, we present the first of two countermeasures designed to protect against balise data integrity attacks: a low-cost and lightweight cryptographic solution that is especially designed to augment the existing balise. Before introducing this device-level countermeasure, we first 
introduce additional technical detail relating to balise telegrams and their coding strategy to provide necessary context. We then discuss  
the technical challenges in addressing the security concerns and the limitations of the current state-of-the-art. Finally, we present the countermeasure and discuss the security assurance it can provide, as well as its efficiency and deployment considerations.
%
Subsequently in Section~\ref{sec:system-level}, we discuss and propose a complimentary system-level countermeasure to secure spot transmission in the railway setting.

\subsection{Telegrams and Coding Strategy}
\label{sec:telegrams}

The Eurobalise telegram is designed with the following key requirements in mind~\cite{EU96}:
\begin{itemize} 
	\item {\bf Safety}: It must be safe in the presence of error events, such as error bursts, random bit errors, bit slips, etc.;
	\item {\bf Large number of information bits}: It must provide at least hundreds of information bits;
	\item {\bf Mixing of format}: For economic reasons, the telegram length should be adjustable to user needs;
	\item {\bf Availability}: Virtually all telegrams should be received correctly.
\end{itemize}

To meet the above requirements, two compatible telegram lengths, 1023 and 341 bits, respectively, are used.\footnote{Note that 1023 is chosen because most good and well-understood cyclic codes have lengths of the form $n = 2^m -1$ for some integer $m$.
Also, 1023 is divisible by 3; this allows the use of a shorter cyclic code (341) together with a longer code (1023) with no mixing cost.}
A distinct and critical feature is that the transmission of a telegram respects cyclic shifts.
That is, the transmission needs not start (or end) at the beginning of a telegram, and thus the detection procedure is completely transparent with respect to cyclic shifts of a telegram.

\begin{figure}[h!]\centering
	\includegraphics[width=0.49\textwidth]{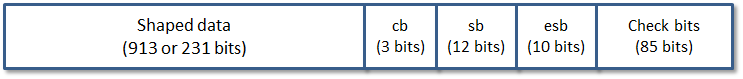}
	\caption{Format of Eurobalise telegram~\cite{Subset36}.}
	\label{fig:telegram_format}
\end{figure}

The telegram format as specified in the Eurobalise standard~\cite{Subset36} is illustrated in Figure~\ref{fig:telegram_format}.
As mentioned, there are two versions: a long format of length $n_l = 1023 = 93 \cdot 11$, and a short format of length $n_s = 341 = 31 \cdot 11$. 
Using the notation from~\cite{Subset36}, the bits of a telegram are denoted by $b_{n-1}, b_{n-2}, \dotsc, b_1, b_0$ (with $n = n_l = 1023$ or $n = n_s = 341$). 
The numbering with descending indices (from left to right) is chosen such that ``left'' and ``right'' conform with Figure~\ref{fig:telegram_format}. 
The order of transmission is from left to right (but need not begin with the leftmost bit $b_{n-1}$).
The format can be described as follows:

\begin{itemize}
	\item {\bf Shaped data} which contains the user data scrambled and shaped during encoding. In the long format, this block consists of 913 bits, while in the short format, the block consists of 231 bits.
	
	\item {\bf Control bits} (cb) where the first bit is the ``inversion bit'', which shall be set to zero.\footnote{This is needed to detect the case when all bits of the telegram have been inverted.} The other two control bits are not currently used and are intended for future format variations. For the present format, these spare bits shall be set to $b_{108}=0$ and $b_{107}=1$.  
	
	\item {\bf Scrambling bits} (sb) are used to store the initial state of a scrambler that operates on the data bits before shaping.
	
	\item {\bf Extra shaping bits} (esb) are used to enforce the shaping constraints on the check bits independent of the scrambling. They are disregarded by	the receiver (except that the shaping constraints are checked).
	
	\item {\bf Check bits} comprise 75 parity bits for error-detecting and 10 bits for synchronization.
\end{itemize}

User data includes a 50-bit header (containing 14-bit balise group identifier allowing for a unique ID of every balise group) followed by multiple packets.
Typical data packets used in a telegram are linking data (distance to the next balise group), movement authority (maximum speed), gradient profile (uphill/downhill), and end of information.






In a nutshell, the encoding scheme for Eurobalises comprises the following three key steps:
\begin{enumerate}
	\item {\bf Scrambling}: This is to ``shape'' (or randomize) the user data (or user bits) to facilitate robust and efficient modulation/demodulation during data transmission, i.e., eliminate long sequences of 0's or 1's. 
	The data scrambling algorithm takes as input the scrambling bits ($sb$) as initial state to derive a scrambling key $S$, which in turn, is fed to a linear feedback shift register (LFSR) along with the user data. We note that such a LFSR-based randomization process is not designed to be cryptographically secure\footnote{In the current Eurobalise encoding scheme, both $sb$ and $S$ are not considered secret values.}, but rather for data scrambling purposes.
	
	\item {\bf Substitution}: In this step, the scrambled bits are partitioned into blocks of 10 bits each. 
	Each such block shall be transformed into an 11-bit word by a substitution table containing 1024 values (in the order of increasing magnitude), which are listed in~\cite{Subset36}. 
	
	\item {\bf Computing check bits}: In the last step, a cyclic code based on Bose-Chaudhuri-Hocquenghem (BCH)~\cite{BR60} is computed to ensure that the telegram is sufficiently resistant to various form of transmission errors. However, as we describe in Section~\ref{sec:localization-state-of-the-art}, such cyclic code is not a cryptographically secure checksum.
\end{enumerate}

On the other hand, the corresponding decoding scheme to be performed by a receiver, i.e., on-board transmission equipment, on the received bits involves the following steps: 
\begin{enumerate}
	\item {\bf Window shifting}: Consider a window of $n+r$ consecutive received bits, where $r=77$ for long format and $r=121$ for short format. If the window has already been shifted over 7500 bits, set $r=n$. Repeat the following checks by shifting the window by 1 if any of them fails:
	\begin{itemize}
		\item Is the parity check satisfied, i.e., are the first $n$ bits (considered as a polynomial) divisible by the relevant generator polynomial? 
		\item Do the $r$ extra bits (rightmost in window) coincide with the first $r$ bits (leftmost in window)?
		\item Are all 11-bit words valid?		
	\end{itemize} 

	\item {\bf Verifying control bits}: This step checks the validity of the inversion bit ($b_{109}$) and the control bits ($b_{108}$ and $b_{107}$). 
	\item {\bf Substitution and descrambling}: In this step, the received bits are partitioned into blocks of 11 bits each and transformed back to blocks of 10 bits. Finally, they are descrambled into the original user bits.
\end{enumerate}

Full details of the Eurobalise encoding and decoding schemes can be found in ~\cite{EU96,Subset36}.

\subsection{Technical Challenges}
\label{sec:challenges}

\subsubsection{Setting}

Eurobalise-compliant telegrams have relatively short sizes, either 1023 or 341 bits, out of which 85 bits are check bits.
These fixed sizes are designed to support various train speed and different types/sizes of balise.
The total number of bits transferred in one train passage is dependent on the data rate, the contact length (physical length of the balise) and the train speed.
Due to the nature of intermittent communication in spot transmission, the block length (of a telegram) is chosen such that the number of bits transferred per train passage is always at least three block length.


Given the above requirement, it is not at all clear how one could embed cryptographic secure authentication code which offers a sufficient level of security (i.e., integrity protection) without expanding the sizes or modifying the format of current telegrams.
Moreover, as required by the Eurobalise standard~\cite{Subset36}, up-link data transmission from a balise to the antenna unit of on-board transmission equipment is performed {\em without handshaking}.
This rules out the possibility of deploying a classic challenge-response authentication protocol widely used in wireless communication, for example those proposed in~\cite{BVS13}, on fixed balises.\footnote{While in principle a challenge-response authentication protocol may be deployable on controlled balises (which support both up-link and down-link communications), controlled balises are designed for communicating variable signal data from a traditional signalling system via the LEU~\cite{Trainguard,Subset36} (see Section~\ref{sec:system-design} for usage of LEU and controlled balises).}

Furthermore, according to the IEEE standard 1474.1 for CBTC~\cite{IEEE1474}, the trackside CBTC equipment reaction (nominal) times should be between 0.07 and 1 sec, while the train-borne CBTC equipment reaction times between 0.07 and 0.75 sec.
This implies that whatever techniques or mechanisms used to protect the integrity of telegrams should be sufficiently {\em lightweight}.

\subsubsection{Legacy}

Eurobalise-compliant telegrams have fixed data format and structure.
Moreover, they have already been widely deployed.
Therefore, ideally any introduction of security mechanism for securing telegrams should not result in major changes to the current specification.
For example, we could easily destroy the algebraic properties embedded in a telegram if we introduce new authentication code into the current telegram structure in a naive way.
Also, any message expansion would not be desirable.
This rules out the use of a standard MAC in a straightforward manner.

\subsubsection{Design Constraints}

While the techniques proposed in~\cite{BJO09,DNS15,DNSL15} could be a sensible starting point, there remain non-trivial technical challenges that need to be addressed.
For example, we may consider adopting the concept of integrity-protected error-correcting code (IP-ECC)~\cite{BJO09}.
More concretely (but informally), we may combine both MAC and CRC for achieving corrupt-resilient MAC of the form
\[
h_k(m) \oplus g_{k'}(m')
\]

where $h$ is a linear UHF family, $g$ is a PRF family, and the $\oplus$ operator is bitwise exclusive-OR.
That is, the check bits of a telegram (see Figure~\ref{fig:telegram_format}) is replaced by IP-ECC as defined above.
However, we are still left with the following unsolved issues:
\begin{itemize}
	\item MAC destroys the cyclic property required in the telegram, and thus, may prevent or slow down the decoding process;
	\item IP-ECC would result in message expansion, since the input ($m'$) to the $g$ function should be made available to the receiver;
	\item Key $k$ for the $h$ function should be secret, however, the key (i.e., polynomial generator) in the telegram is publicly known;
\end{itemize}

The above challenges seem non-trivial and may require techniques beyond the realm of coding.

\subsection{Our Integrity Protection Approach}

\subsubsection{Assumption}
Henceforth, we use {\em encoder} to denote the Eurobalise encoding scheme and {\em decoder} to denote the corresponding decoding scheme as described in Section~\ref{sec:telegrams}.
In our approach, we treat the encoder and decoder as ``black boxes''. 
That is, we use the encoder/decoder as it is as a building block to which our technique is applied to realize our solution.
No modification on the design of the encoder/decoder is required.

\subsubsection{Key Idea}

The scrambling bits ($sb$) of a telegram is unstructured and random-looking.
Hence, these seem to be a suitable avenue for embedding an authentication tag.
Intuitively, we bind user data to scrambling bits ($sb$) such that any change to the user data would result in different scrambling bits.
However, one issue with such an approach is that the size of $sb$ is too small, i.e., only 12 bits.
We, therefore, also make use of the corresponding 32-bit scrambling key ($S$), which is taken as input by the LFSR circuit of the encoder, as part of the authenticity/integrity check.
More concretely, we employ a technique that binds user data to the scrambling bits ($sb$) and the LFSR scrambling key ($S$) of a telegram, in a way that allows the reader (on-board transmission equipment) to verify the authenticity and integrity of the telegram.
This way, $sb$ can serve as an authentication tag without explicitly expanding the size of the telegram.
Recall that $S$ is derived from $sb$ (in the current encoder).
In addition, we require that both $sb$ and $S$ are constructed based on a pair of distinct and secret cryptographic keys ($k_0, k_1$), respectively.
In other words, our technique ensures that only an authorized party with knowledge of the keys is able to generate and verify the correct $sb$ and $S$ values for some user data.
Since the length of $S$ is still relatively smaller than standard sizes used for cryptographic primitives (e.g., 80 or 128 bits), we require that the key pair ($k_0, k_1$) must be unique for each balise and no tag verification queries are allowed.

\subsubsection{Implication}

Our technique is designed such that any modification to user data would lead to incorrect descrambling of the associated shaped data.
In order to forge valid shaped data, an adversary is forced to first construct both valid $sb$ and $S$. 
This, in turn, requires knowledge of both keys ($k_0, k_1$).
Note that here, $sb$ is analogous to a traditional MAC tag; hence, it needs not to be kept secret and it should not leak information about the key on which it was based.
Similarly, $S$ is generated such that it leaks no information about the associated key from which it was constructed.

\subsection{Construction of Authentication Tag}

In what follows, we use $\mathsf{F}$ to denote a key derivation function, $\mathsf{MAC}$ a message authentication code scheme, and $\mathsf{PRF}$ a pseudorandom function.

\subsubsection{Key Generation}

Let $mk$ a master key and $id$ a unique balise identifier.\footnote{We assume that the balise identifier can be extracted from a track-map database accessible by the on-board transmission module.}
Our key generation algorithm then computes the required authentication key pair for each balise by setting
\[
k_0 := \mathsf{F}(mk, id, \text{`0'}),\;\; k_1 := \mathsf{F}(mk, id, \text{`1'})
\]
where $|k_0|$ and $|k_1|$ can be 128 bits, for example.\footnote{Different key sizes may be used but may result in different security vs.\ efficiency trade-offs.}

\subsubsection{Tag Generation}

\begin{figure}[h!]\centering
	\includegraphics[width=0.3\textwidth]{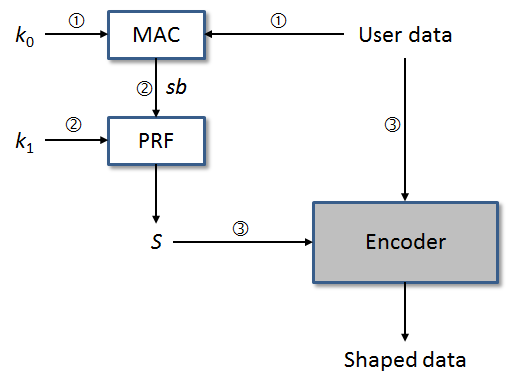}
	\caption{Generation of authentication tag $sb$.}
	\label{fig:gen_tag}
\end{figure}

Our authentication tag generation algorithm performs the following steps:
\begin{enumerate}
	\item Take as input the user data $U$ and the key $k_0$, set $sb$ to be the first 12 bits of $\mathsf{MAC}(k_0, U)$;
	\item Take as input the key $k_1$ and the scrambling bits $sb$, set $S$ to be the first 32 bits of $\mathsf{PRF}(k_1, sb)$;
	\item Feed $S$ and the user data $U$ into the encoder (as the initial state of the LFSR circuit).	
\end{enumerate}
The output authentication tag here is $sb$.
Our tag generation algorithm is illustrated in Figure~\ref{fig:gen_tag}.
Note that while $sb$ is explicitly specified in a telegram, $S$ is not, but must be computed from $sb$.

\subsubsection{Tag Verification}

Given an authentication tag $sb$, our tag verification algorithm performs the following steps:
\begin{enumerate}
	\item Take as input the key $k_1$, set $S'$ to be the first 32 bits of $\mathsf{PRF}(k_1, sb)$;
	\item Feed $S'$ into the decoder to recover user data $U'$;
	\item Take as input the key $k_0$ and decoded user data $U'$, set $sb'$ to be the first 12 bits of $\mathsf{MAC}(k_0,U')$;
	\item Check if $sb = sb'$, if so, verification passes.
\end{enumerate}

\begin{figure}[h!]\centering
	\includegraphics[width=0.47\textwidth]{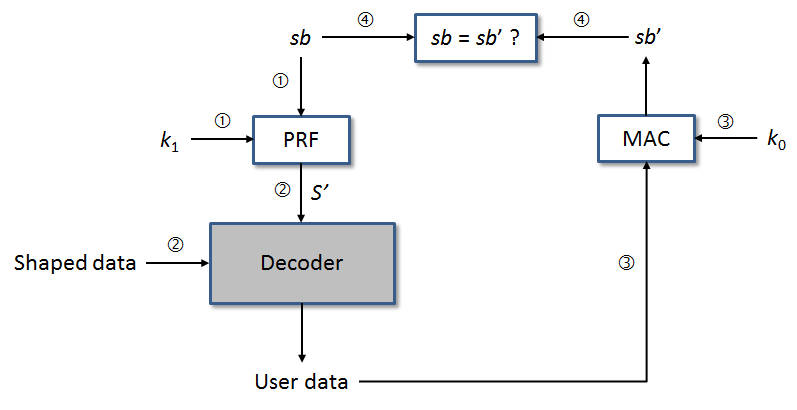}
	\caption{Verification of authentication tag.}
	\label{fig:ver_tag}
\end{figure}

It is important to note that we treat the current encoder and decoder as black boxes in our approach, in the sense that we did not make modification to existing the LFSR-based scrambling and check bits generation algorithms.
Instead, we merely define how $sb$ and $S$ are computed before the actual encoding process takes place, and how these values are verified before decoding.

\subsubsection{Key Management}

Standard key management techniques can be applied to derive and protect the master key $mk$.
For example, each railway line may use a distinct $mk$ derived from a root key owned by the relevant railway operator.
We assume that all keys, including $mk$, $k_0$, and $k_1$ are not updated regularly, unless is necessary due to security compromise, for example.
Hence, it is critical that the root key and each master key are appropriately protected, for example, through a tamper-resistant storage system.
We envision that one possible way to update the keys ($k_i$ for $i \in \{0,1\}$) associated with a specific balise is by setting
\[
k_i := \mathsf{F}(mk, id, ver, i)
\]
where $ver$ denotes a version number.
Such a key diversification technique is commonly used in public transportation systems based on contactless smart card payment technology.
For example, EZ-Link cards deployed in Singapore are compliant to the CEPAS standards~\cite{KLG13,ss518}, which in turn, require that each terminal or card reader has embedded multiple distinct master keys in a secure access module.
This is to avoid reissuance of a master key in the event of key exposure during the life-time of the terminal or reader.
Formal security analysis of key derivation and diversification techniques based on bootstrapped master keys has been well-studied, see for example~\cite{BP10}.

\subsection{Security}

Recall that our security goal is to ensure that the adversary is unable to forge or construct valid authentication tag ($sb$) and LFSR scrambling key ($S$) without the knowledge of keys ($k_0, k_1$), respectively. 
Since both $sb$ and $S$ are bound to the corresponding user data, any modification to the user data would be detected through the tag verification algorithm described before.

We now (informally) examine the security guarantee provided by $sb$ and $S$.
The former is essentially a truncated MAC, while the latter is a truncated output of a PRF.

\subsubsection{Truncated MACs}

There are two main classes of attack on a MAC scheme, namely key recovery and forgery attacks.
In the former, an attacker attempts to discover the secret key used to compute the MACs, while in the latter, an attacker attempts to determine the correct MAC for a message without knowledge of the corresponding secret key.
When assessing the security of a MAC scheme, we typically quantify the resources needed for the attack in terms of the required numbers of known message/MAC pairs, chosen message/MAC pairs, and on-line MAC verifications.
Mitchell~\cite{Mitchell03} showed that truncated MACs could be vulnerable to serious attacks should the following two assumptions hold: (i) the attacker is allowed to submit messages and accompanying MACs, and determine whether or not the MAC is correct; and (ii) the attacker is able to determine the degree of truncation of the MAC, i.e., the MAC length.
However, these two assumptions do not hold in our context since our approach does not allow any tag verification queries to be made by the attacker, and the MAC length is always fixed, i.e., 12 bits.
In other words, we restrict the probability of a successful MAC forgery to $1/2^{12}$.

A more recent study by Ga\v{z}i, Pietrzak, and Tessaro~\cite{GPT15} provides a proof of the security inherent in truncated CBC-MACs (TCBC).  
Let $n$ be a block size of TCBC and $r$ a relatively small (truncated) output length.
They proved that no polynomial-time adversary making $q$ queries of length at most $l < 2^{n/4}$ to TCBC using a random permutation can distinguish it from a random function (returning random output for each distinct message), except with a distinguishing gap of
\[
\epsilon(q) = \mathcal{O} \left(\frac{lq^2}{2^n} + \frac{q(q+l)}{2^{n-r}}\right).
\]
It is easy to see that the above security bound is reasonable for small $r$, as long as $q$ is kept fairly small ($< 2^r$) and $n$ is sufficiently large, for example $n=128$.

Moreover, as pointed out in~\cite{GPT15}, there is a folklore belief that given a secure MAC, truncating its output may actually increase its security by hindering collision detection.
However, this has never been formally verified at the time of writing.

\subsubsection{Pseudorandom Numbers}

It is well-known that if $\mathsf{PRF}$ is a pseudorandom function, and given a random $x$ value, $\mathsf{PRF}(x)$ can be served as a one-time pad outputting an unpredictable, random value.
In our setting, we define $S = \mathsf{PRF}(k_1, sb)$, where $sb$ (truncated MAC) is considered random and key $k_1$ is used to ensure that $S$ can only be correctly computed by an authorized entity with knowledge of the key.
Moreover, since $\mathsf{PRF}$ is a one-way function, even if the attacker has knowledge of $S$, no information about $k_1$ can be learned.
To construct a valid $S'$ for user data $U'$, the attacker must first forge a valid $sb'$ corresponding to $U'$, and then predict $S'$ corresponding to the forged $sb'$.
This is computationally infeasible if $\mathsf{MAC}$ and $\mathsf{PRF}$ are both cryptographically secure.

\subsection{Efficiency}

We tested our algorithms implemented in Python 3.4.3, on Inter Core i3-2120 CPU at 3.30 GHz and 8GB of RAM, and running with Ubuntu 14.04.4 LTS.					
Particularly, we compared the computation time required to calculate $S$ in both cases: current Eurobalise encoder/decoder (without security protection), and our cryptographically integrity-enhanced approach.	
Our experimental results (taken over 10 iterations) are summarized in Table~\ref{tab:computation_time}.

\begin{table}[h] \small 
	\begin{center}	
		\caption{Computation time (in ms) for calculating the $S$ value.} 
		\label{tab:computation_time}
		\begin{tabular}{lcc}
			\toprule
			& \multicolumn{1}{c}{Encoder} & \multicolumn{1}{c}{Decoder} \\ 
			\midrule
			Current Eurobalise & 0.0047 & 0.0016 \\
			Our approach & 0.0274 & 0.0145 \\
			\bottomrule
		\end{tabular}
	\end{center}
\end{table}

Our evaluation shows that the additional computational cost incurred for ensuring data integrity protection is negligible, approximately in the range of between 0.013 and 0.023 ms.
Note here that the encoder is typically executed offline (during initialization or deployment of balises) using a balise encoder programming tool~\cite{Atlas,TPG-Eurobalise}; while the decoder is run by an on-board transmission equipment with reasonably powerful computing resources.
Also, we do not evaluate the latency caused by our algorithms because we did not modify the format and size of the Eurobalise telegram and its parameters.
Hence, in practice, we do not foresee any noticeable performance impact to the actual deployment of a spot transmission system.

\subsection{Deployment}

We reiterate that our technique of embedding an authentication tag does not require any changes to the existing standardized encoding/decoding schemes.
What is needed is to update the software of a balise programming tool (e.g., \cite{Atlas,TPG-Eurobalise} such that the scrambling bits ($sb$) and the scrambling key ($S$) are generated as specified before.
Moreover, the on-board transmission equipment of a train needs to be updated with our tag verification algorithm and given access to the required cryptographic keys in a secure manner.

\section{Integrity Protection: System Level}
\label{sec:system-level}
In this section, we discuss the second of our two countermeasures: a system-level approach considering information from multiple sources to determine whether the telegram received from a balise is trustworthy.

\subsection{Importance of System-level Logic}
Recall from Section~\ref{sec:threatmodel} that we consider two types of attacks on balise spot transmission systems: tampering attacks, and cloning attacks. The device-level countermeasure presented above secures the communication interface between a balise and the train's onboard receiver. The presence of an authentication tag addresses tampering attacks---an adversary who does not possess the necessary keys will be unable to program a valid (modified) telegram on a balise. 

While this authentication scheme provides a strong layer of protection on its own, there are two important areas where security can be further enhanced. First, although a train receiving data from a compromised balise would be able to determine whether or not the data were modified, this device-level countermeasure on its own does not address cloning attacks. An attacker could copy a valid telegram onto multiple balises. A broader system perspective, considering multiple balises and the manner in which they are deployed and used by passing trains, is necessary to cover both types of attacks. Furthermore, a train relying on data from balises for a real-time control process such as speed control needs to know more than simply whether balise data is trusted or not: it needs a response strategy to address situations where data is missing or invalid. The system-level countermeasure we present in the remainder of this section seeks to address these gaps. To provide a specific system use case, we use the train automatic stop control scenario introduced in Section~\ref{sec:usecase}. 



\subsection{Improving Security from System-level Insight} 
The design of our countermeasure is based on four key observations about stop control for automated trains, which we discuss in the paragraphs below. 

\begin{figure*}[t] 
	\centering
	\subfigure[Block diagram of the train stop control model with countermeasure logic.]{
		\includegraphics[width=0.65\linewidth]{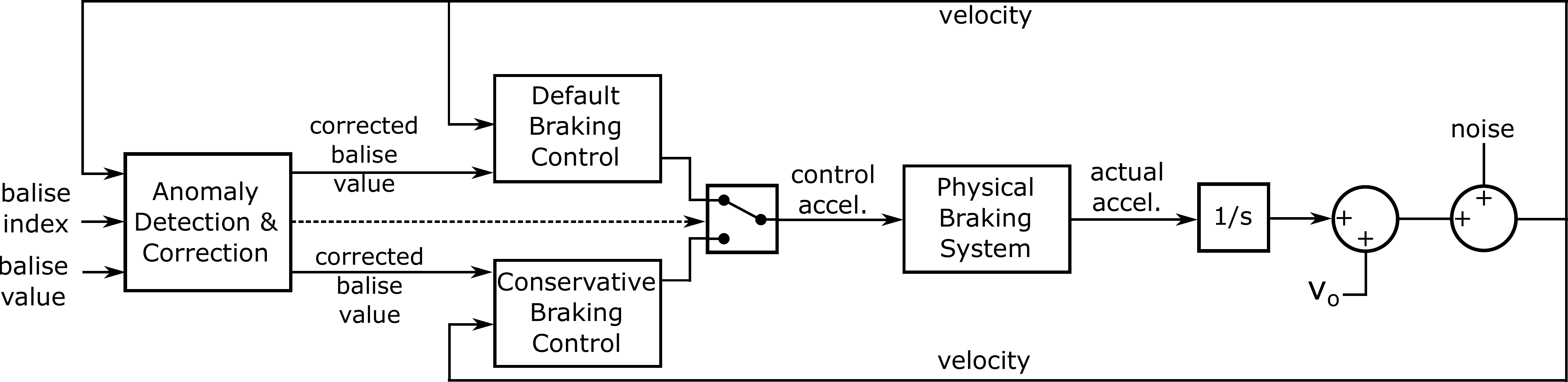}
	\label{fig:blockdiagramcounter}
	}
	\subfigure[Details of conservative controller]{
	\includegraphics[width=0.27\linewidth]{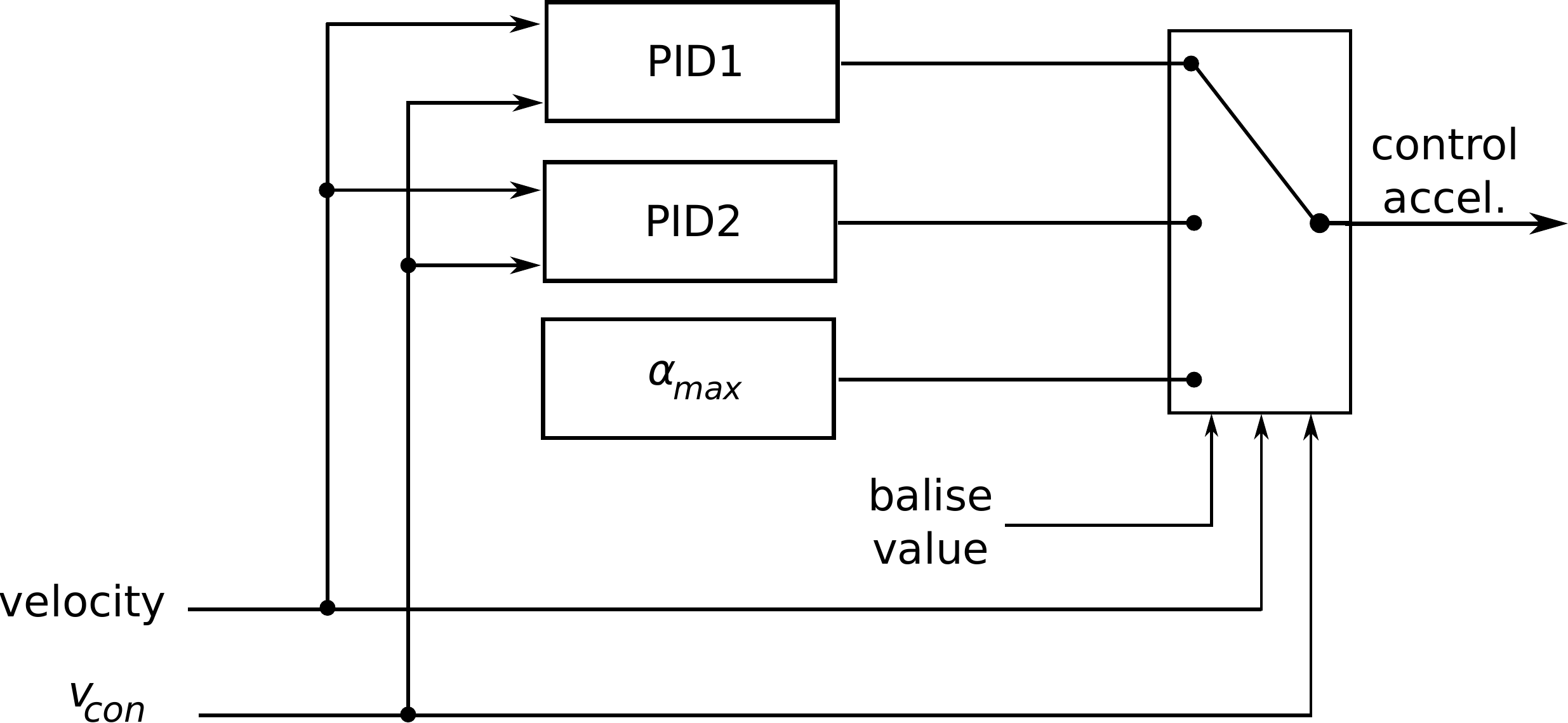} 
	\label{fig:ccdiagram}
	}
	\caption{Resilient train stop control logic with anomaly detection \& correction and conservative braking controller.}
	\label{fig:systemcounter}
	\vspace{-3mm}
\end{figure*}

\subsubsection*{{\bf Observation 1}}
{\itshape Trains can leverage the trustworthy and static physical configuration of a rail system.} In a railway environment, the distance between stations and the position of balises are fixed once the system is built and put into operation. It is a common practice to load such physical-world information into a train's on-board computers (e.g., track maps discussed in~\ref{sec:localization-state-of-the-art}) and a braking controller can use that trustworthy information to make decisions. Note that our threat model in Section~\ref{sec:threatmodel} assumes the attacker cannot change the physical setup, e.g., moving a balise to a different position. Balises have detailed installation requirements governing orientation and distance from the track~\cite{Subset36}, therefore a physical movement attack would be 
time-consuming and easier to detect than a re-programming attack.

\subsubsection*{{\bf Observation 2}}
{\itshape Trains can leverage less accurate but more trustworthy sensing data.} A train can infer its location using multiple onboard sensors, such as a tachometer or Doppler radar sensor. The location estimate from onboard sensors is not as accurate as the value obtained directly from balises. For example, estimates based on wheel rotation can be inaccurate when the wheels slip on the rails.
As a result, the onboard estimation error accumulates with the distance that a train travels from the last reference point. 
Fortunately, with modern information fusion techniques, 
the onboard system is able to provide well-bounded accuracy based on these imperfect sensors. 
For ETCS, the minimum accuracy requirement for on-board train position measurement is $\pm(5m+5\%d)$ where $d$ is the distance travelled since the last reference point~\cite{ETCSsub041}. In practice, the performance 
can be better: for example an error less than $\pm 20m$ for every $1000m$ travelled ($\pm2\%d$~\cite{Malvezzi}). 
Under our threat model, we consider the on-board location estimate to be more trustworthy than location information received from balises, however it is less accurate.

\subsubsection*{{\bf Observation 3}}
{\itshape Trains can reduce stopping error by sacrificing performance}
As a train approaches a station, it will ideally stop with high accuracy, i.e., on or close enough to the target point (balise $B_m$ in our model) for passengers to alight, and it should complete the process of slowing and stopping in the shortest time possible, subject to operating constraints (e.g., maximum braking rate). In normal situations, the position references received by fixed balises help the train optimize its braking curve for these two objectives. As we will show in Section~\ref{sec:evaluation}, tampering or cloning attacks can make the train overshoot or undershoot the target stop point by a wide margin. However, 
even in the case where a train cannot receive any trustworthy position input from fixed balises during the approach, it 
can still stop close to the right position by travelling at a low speed and applying the brake when it passes the controlled balise at the stop point which interfaces with the platform screen doors. 
If the speed is low enough, the train should be able to stop within a short distance from the correct position. The drawback of this conservative strategy is that the train will take longer to stop.



\subsubsection*{{\bf Observation 4}}
{\itshape Trains can use trusted information to verify information from untrusted sources.} As discussed above, knowledge of the physical configuration of the train system (e.g., station, balise locations) and the position estimate derived from on-board sensors are considered to be trustworthy information. This information can be used to cross check the input from less trustworthy sources: namely, the balise telegrams.
Once the information from balises is deemed trustworthy, it can be used to improve performance.

\subsection{Our Resilient Control Approach}
Based on the four system-level observations above, we have developed a software-only countermeasure for dealing with attacks on balise telegrams in train automatic stop control systems. As with the device-level countermeasure in Section~\ref{sec:device-level}, this scheme does not require changes to the legacy balise infrastructure that is in use today: it simply requires new braking logic in the trains themselves. 

To describe our countermeasure in detail we introduce some additional notation: let $p$ denote the train's position as it approaches a station; let $p_{est}$ denote the train's onboard position estimate, which has an associated position error $\delta$; let $v_{con}$ denote a reduced speed setting. 
At a high level, our resilient controller uses two parallel control modules---the original controller (e.g., PID, HOA, or GOA~\cite{Chen2013balise}), and a new conservative controller---which are selected based on an anomaly detection and correction module, which includes integrity checking via the scheme in Section~\ref{sec:device-level}. This control architecture is shown in Figure~\ref{fig:blockdiagramcounter}. The logic is described in Algorithm~\ref{alg:robust}. 
Note that this controller is effective against attacks rendering balise telegrams unavailable 
\cite{Temple2017} as well as against the tampering and cloning attacks discussed in this paper. 

By default a train will operate using the normal controller: in this case the heuristic online algorithm~\cite{Chen2013balise}. If the Anomaly Detection \& Correction module detects that there is a balise missing, by using it's knowledge about the physical world setup, its onboard estimated position $p_{est}$ in conjunction with whether or not data is received, (Algorithm~\ref{alg:bmissing})
it will fall back to a conservative controller (Observation 3). 
Such a detection is possible, thanks to the static physical-world setup information (Observation 1) and the less accurate but more trustworthy local sensing data (Observation 2).

Internally, as shown in Figure~\ref{fig:ccdiagram}, the conservative controller is composed of 2 PID controllers which act depending on the velocity of the train. 
The first PID controller, \textit{PID1}, is tuned to be more aggressive with the main objective of quickly bringing down the speed of the train to $v_{con}$. The second PID controller \textit{PID2} adopts less aggressive parameters to stably maintain the speed of the train at $v_{con}$ till it reaches the stop marker. In addition to the two PIDs, there is a multiplexer block to switch between these two PIDs as well as ($\alpha_{max}$).
At first \textit{PID1} is selected, and the multiplexer only switches to \textit{PID2} when the velocity of train reaches $v_{con}$. Once the stop marker is detected, the multiplexer will switch to maximum braking of $\alpha_{max}$. There is a need to select between multiple PIDs at different velocity stages due to conflicting requirements in terms timing and stability. To explain this rationale, assume there is only \textit{PID1} in the design. \textit{PID1} promptly helps lower down the speed of the train to $v_{con}$ but because of its aggressively-tuned nature, maintaining the speed at $v_{con}$ becomes a challenge. Wide undershooting can make the velocity 0, stopping the train ahead of the stop marker. On the other hand, if there is only \textit{PID2} in the design, the slower response of this controller takes the train much longer time to reach $v_{con}$. By the time the velocity reaches $v_{con}$, there can be cases that when the stop marker is detected, the velocity of the train is much higher than $v_{con}$ and switching to $\alpha_{max}$ stil fails to to make it stop within the allowable stopping error. Through experiments we obtain the parameters for \textit{PID1} $K_p=0.8423, K_i=0.0648, K_d=0.4082$  as and for \textit{PID2} as $K_p=0.0377, K_i=0.0002, K_d=0.2205$.

\begin{algorithm}[t] \small 
	\caption{\label{alg:robust} {\bf Resilient Train Stop Control}}
	\begin{algorithmic}[1]
		\WHILE{true}
		\IF{Running controller is DEFAULT}
		\IF{Balise-Missing() returns TRUE}
		\STATE Switch from DEFAULT to CONSERVATIVE;
		\ENDIF
		\ELSIF{Encountering a new balise}
		\IF{Derive-Trustworthy-Info() returns TRUE}
		\STATE Update the trustworthy information;
		\ENDIF
		\ENDIF
		\ENDWHILE
	\end{algorithmic}
\end{algorithm}
\begin{algorithm}[t] \small 
	\caption{\label{alg:bmissing} {\bf Balise-Missing()}}
	\begin{algorithmic}[1]
		\STATEx // {\bf Train is provided with known balise locations before operation. In this paper, the known locations are assumed to be [-100, -64, -36, -16, -4]}
		\STATE update $p_{est}$ and $\delta$ based on trustworthy information;
		\IF{there is a balise location $loc_i$ such that $abs(p_{est}) < abs(loc_i) - \delta$ {\bf or} $abs(p_{est}) < abs(loc_{i+1}) + \delta$ {\bf while} no position reference of $loc_i$ has been received}
		\STATE {\bf return} TRUE;
		\ENDIF
		\STATE {\bf return} FALSE;
	\end{algorithmic}
\end{algorithm}

\begin{algorithm}[!t] \small 
	\caption{\label{alg:dtrustworthy} {\bf Derive-Trustworthy-Info(), with $(p_{est}, \delta)$ being the estimate so far.}}
	\begin{algorithmic}[1]
		\IF{AuthenticateBalise()=TRUE} // could be correct balise or replay attack
        \IF{reported position $loc^\prime_i$ satisfies $|p_{est} - loc^\prime_i| < \delta$}
        \STATE {\bf return} $(loc^\prime_i, 0)$ as new $(p_{est}, \delta)$; // otherwise estimate remains
        \ENDIF
        \ELSIF{AuthenticateBalise()=FALSE} // balise has been tampered with. Try to recover useful information
        		\IF{there is a unique known balise position $loc_i$ such that $|p_{est} - loc_i| < \delta$}
		\STATE {\bf return} $(loc_i, 0)$ as new $(p_{est}, \delta)$;
        	\ELSE{} $\qquad$ //~there are multiple known balise positions $loc_i \in B$ such that $|p_{est} - loc_i| < \delta$ 
            \STATE set $r \leftarrow \{ local = p_{est}, candidates = B \}$;
		\STATE append $r$ into $R$, an unverified record list;
        \IF{there are multiple records $r_1, \ldots r_k$ in $R$}
        \FOR{$i$ in $1, \ldots, k-1$}
		\STATE $d \leftarrow r_k.local - r_i.local$;
		\STATE  compute distance between all combinations of $r_i.candidates$ and $r_k.candidates$;
		\IF{there is only one pair of candidates $(loc_l, loc_m)$ that has distance of $d$}
		\STATE {\bf return} $(loc_m, 0)$ as new $(p_{est}, \delta)$;
		\ENDIF
		\ENDFOR	
        \ENDIF
        \ENDIF 
        \ENDIF
	\end{algorithmic}
\end{algorithm}

\section{Evaluation} 
\label{sec:evaluation}
From Sections V and VI we see that balise transmission can be secured at both the device and system levels. In this section, we simulate a train stopping at a station to illustrate the negative impact of tampering and cloning attacks, as well as the benefits derived from implementing our countermeasures. We start by describing simulation parameters and the normal (no attack) scenario to illustrate how automated trains use balise input during a stop.

\subsection{Normal Braking Scenario}
As discussed in Section~\ref{sec:usecase}, this scenario concerns an automated train approaching a station and using position data from sequence of balises to adjust its brake force and stop in alignment with the platform screen doors. This model is based on a particular system used in Asia~\cite{Chen2013balise}, however we are aware of other train lines employing a similar operating principle with differences in the balise layout and braking control. 

We consider a train approaching a station with balises located at positions 
$\mathbf{loc}=[-100, -64, -36, -16, -4, 0]$.
The first five balises are fixed, while the final one is controlled. There are several parameters that characterize the train stop scenario. We assume an initial position of $-100m$, initial velocity of $10m/s$, maximum deceleration of $\alpha_{max}=-1 m/s^2$, and an allowable stopping error of $\gamma=0.3m$. In our analysis we vary the train brake parameters---the time delay ($T_d$) and time constant ($T_p$)---to capture the impact of control disturbances and variations between different trains. We consider default values $T_d=0.6$ and $T_p=0.4$. 
%

\begin{figure}[t] 
	\centering
	\subfigure[Train acceleration with no attacks.]{
		\includegraphics[width=0.83\linewidth]{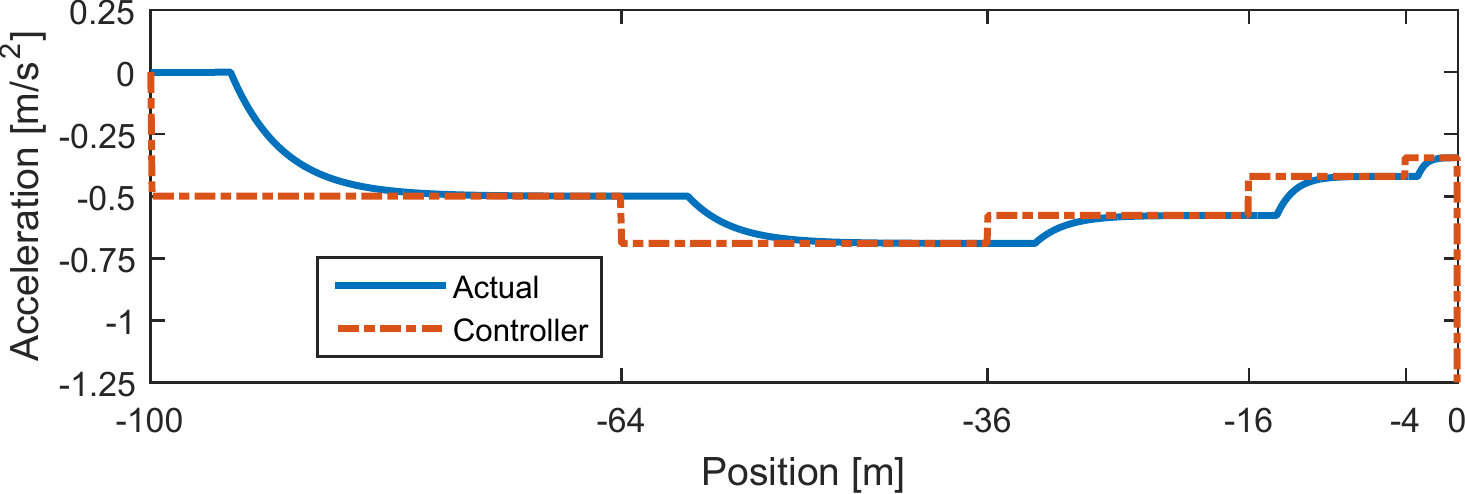} 
		\label{fig:basecase1}
	}
	\subfigure[Train speed with no attacks.]{
		\includegraphics[width=0.8\linewidth]{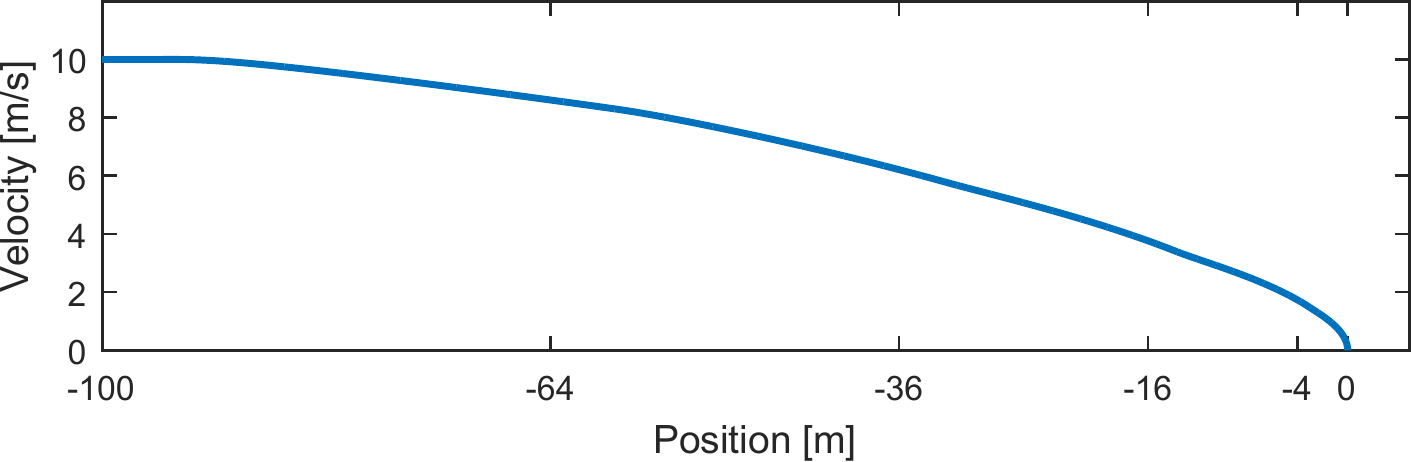} 
		\label{fig:basecase2}
	}
	\caption{Train acceleration/speed with HOA controller and default parameters. }
	\label{fig:accelerationplots}
	\vspace{-3mm}
\end{figure}

To illustrate the train's braking performance in the normal (no attack) case, 
Figure~\ref{fig:basecase1} shows the train's acceleration curve 
under the default simulation setting described above. Observe that at each balise's location the train's brake controller acceleration changes and the actual acceleration follows the change based on parameters $T_d$ and $T_p$. Figure~\ref{fig:basecase2} shows the resulting train speed curve, which achieves a smooth and accurate stop. 

If an attacker carries out a tampering or cloning attack, the timing of brake actions or the degree of braking may be altered, leading to a stop outside of the allowable range of $\pm 0.3m$ from the target. If a train stops too far before or after the target it needs to correct its position. For an unattended train (the highest grade of automation, with no driver onboard) this can be done with a process called jogging, where the train makes a series of incremental movements ($<1m$) until it reaches the target. Alternatively, a human operator may intervene. In either case, the result will be a delay in the train network: the train affected by the attack will take more time at the station, thereby delaying trains behind it. 

\begin{figure}[t] 
	\centering
	\subfigure[Train acceleration with low value attack.]{
		\includegraphics[width=0.8\linewidth]{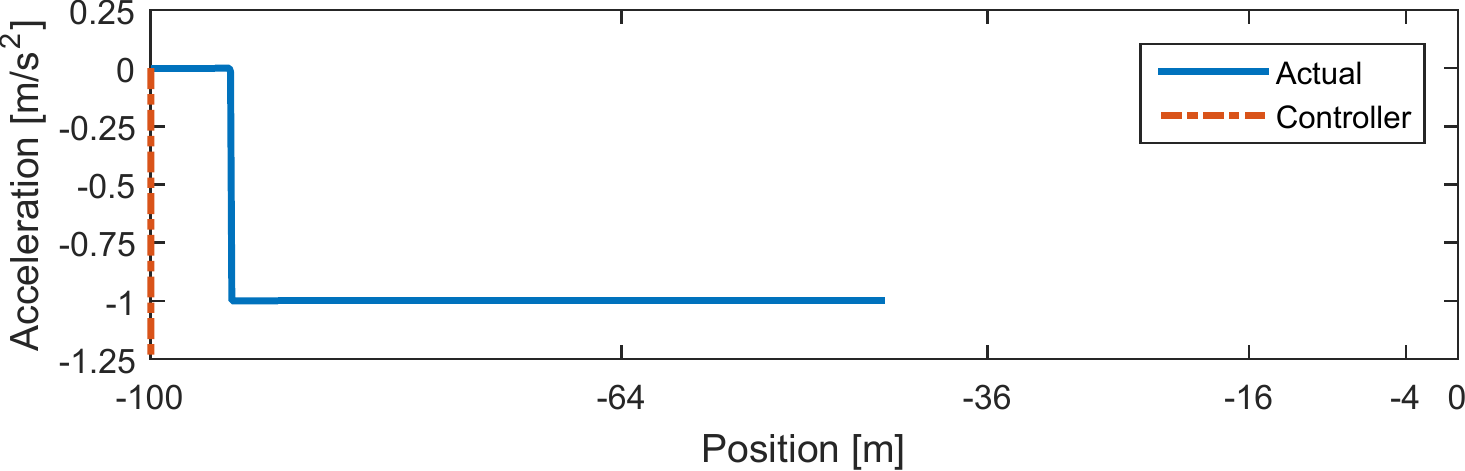} 
		\label{fig:low1}
	}
	\subfigure[Train speed with low value attack.]{
		\includegraphics[width=0.8\linewidth]{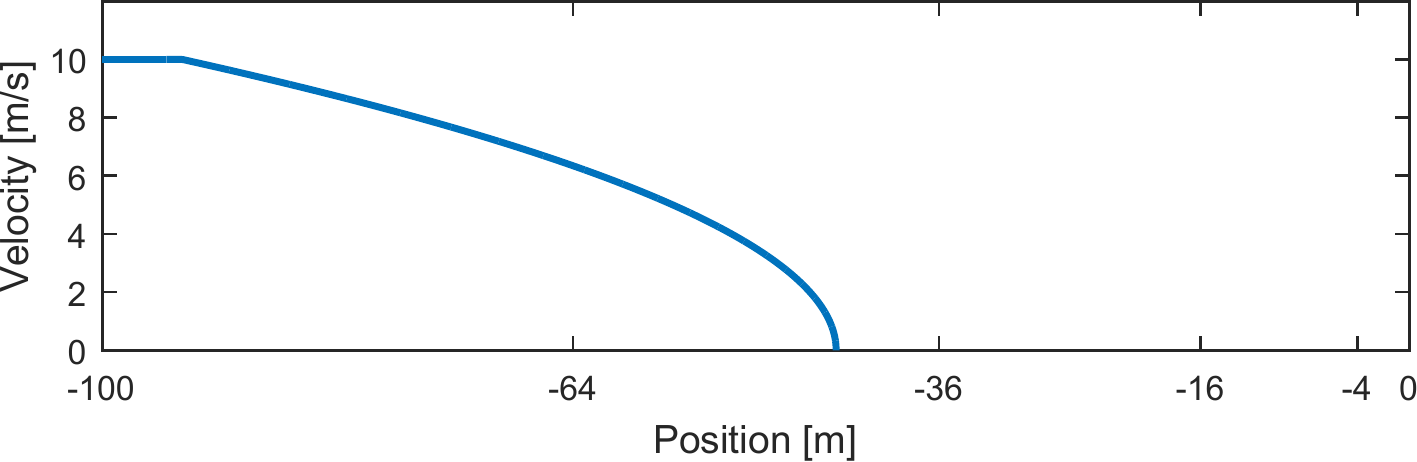} 
		\label{fig:low2}
	}
	\caption{Train acceleration/speed with $B_1$ altered to a low value (-4). }
	\label{fig:lowattack}
	\vspace{-3mm}
\end{figure}
\subsection{Tampering Attack}
\subsubsection*{{\bf Attack Impact}}
An attacker wishing to disrupt train operations could tamper with balises deployed near train stations. 
Intuitively, the further a train can be made to stop away from the target point, the greater the disruption. One potential data integrity attack is re-programming the first (earliest) balise in a stopping sequence to make the train think it has already reached, or is about to reach, the stop point. Using the TASC model, we set $loc^\prime_1=loc_5=-1$ to make the train think it is about to reach the target\footnote{Depending on the system implementation, it may not be possible for a fixed balise such as $B_1$ to impersonate a controlled balise (i.e., $B_6$).}.

Figures~\ref{fig:low1} and~\ref{fig:low2} show the acceleration and speed profiles of the train as it proceeds toward the stopping point (location $0$). The high speed of the train combined with the sudden (perceived) need to stop results in the acceleration controller saturating (dashed line is overlaid with the y axis) and the train slowing at its maximum rate of $-1m/s^2$. As a result the train stops 43.9m before the correct position, and the passengers would be stranded until the train was able to adjust its location. The adjustment could be done via human intervention or by an automatic routine to slowly and incrementally move the train forward (referred to as jogging).

\subsubsection*{{\bf Countermeasure Effectiveness}}
\begin{figure}[t]
	\centering 
	\includegraphics[width=0.83\linewidth]{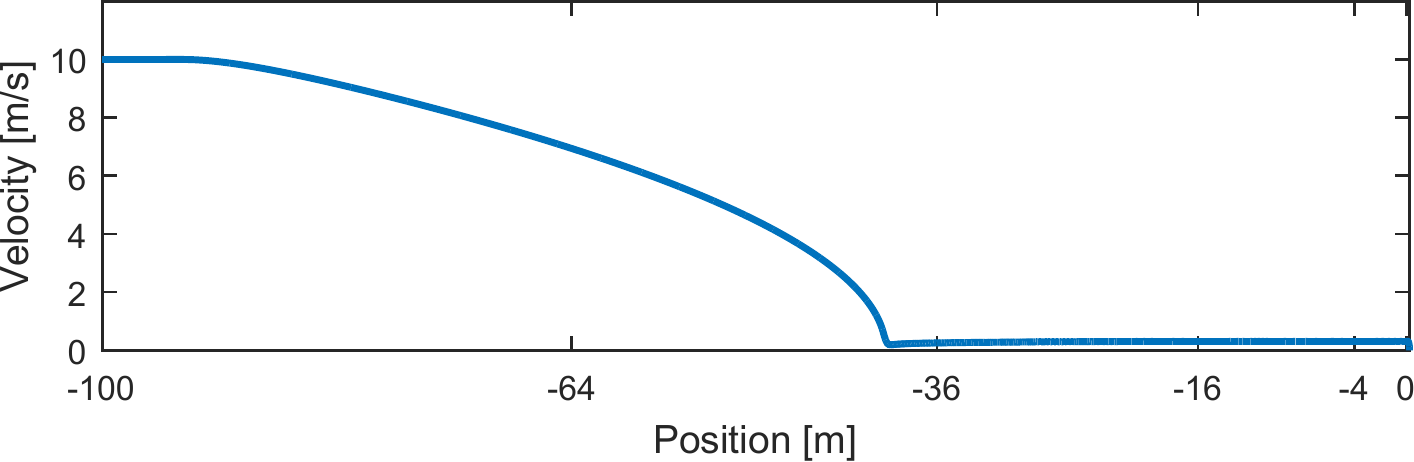} 
	\caption{Speed profile, response to Tampering attack, initial $p_{est}=-80$.}
	\label{fig:ccexample}
	\vspace{-3mm}
\end{figure}

Consider the attack shown in Figure~\ref{fig:lowattack}, where the first balise in the stop sequence is tampered with to report a low value (in this case $-1m$), causing the train to stop early. Implementing the authentication countermeasure in Section~\ref{sec:device-level} renders the attack infeasible, however if the train simply ignores input from balise $B_1$, the result is identical to an availability attack on that balise, which has been shown to cause an overshoot of $1.3m$ beyond the stop point~\cite{Temple2017}. This error, while much smaller than the original $-43.9m$ distance error, is still outside of the allowable $\pm 0.3m$ range and would therefore cause a delay. 
Adding additional system-level logic to respond gracefully, as described in Algorithms~\ref{alg:robust}--\ref{alg:dtrustworthy}, allows the train to infer (from its onboard position estimate) that it has passed balise $B_1$, and begin braking in an appropriate manner. The result is a smooth stop with an allowable stopping distance error of $0.15m$ if the initial $p_{est}=-120$, or $0.23m$ if the starting position estimate is $p_{est}=-80$. In the latter case, the train believes it has missed the first balise and switches to the conservative braking controller; this results in a swift deceleration followed by a slow approach to the target point, as shown in Figure~\ref{fig:ccexample}.

\begin{figure*}[t] 
	\centering
	\subfigure[Train acceleration with duplicate attack (full brake).]{
		\includegraphics[width=0.4\linewidth]{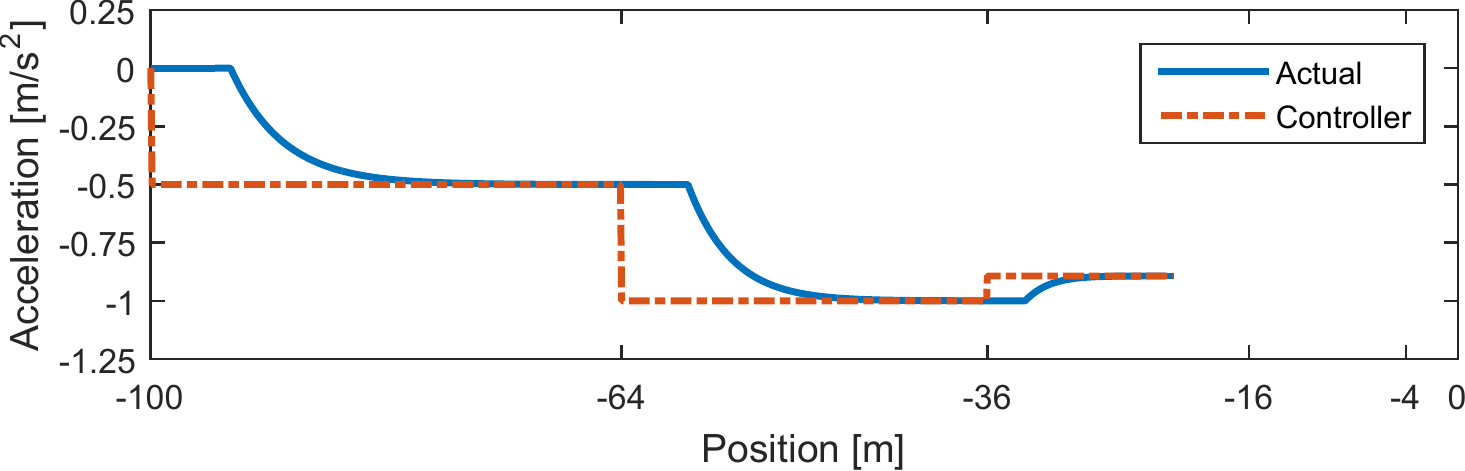}
		\label{fig:same1a}
	}
	\subfigure[Train speed with duplicate attack (full brake).]{
		\includegraphics[width=0.4\linewidth]{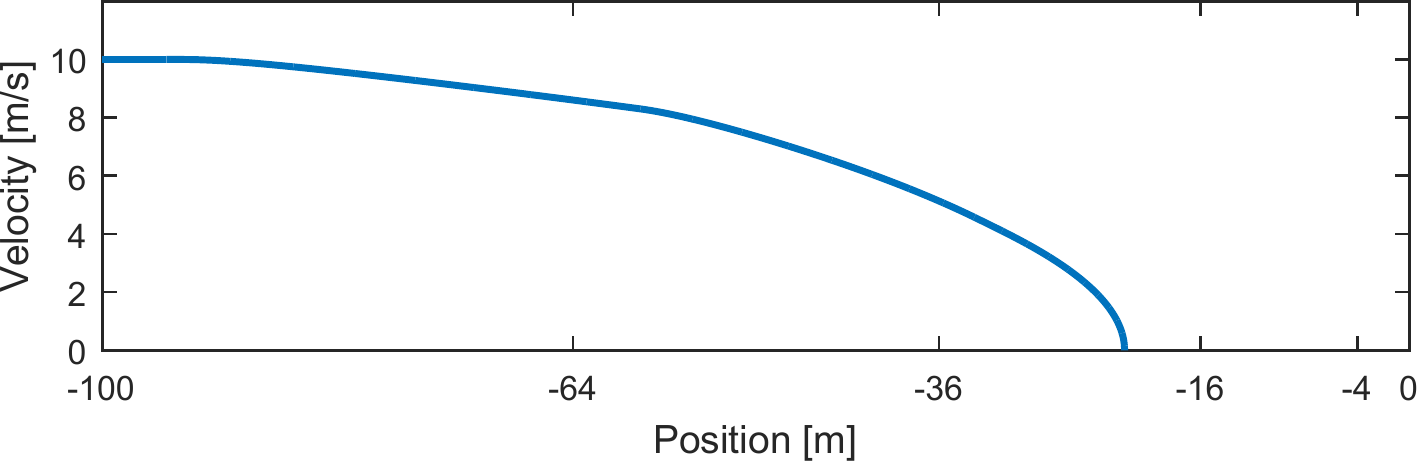}
		\label{fig:same1v}
	}
	\subfigure[Train acceleration with duplicate attack (maintain accel.).]{
		\includegraphics[width=0.4\linewidth]{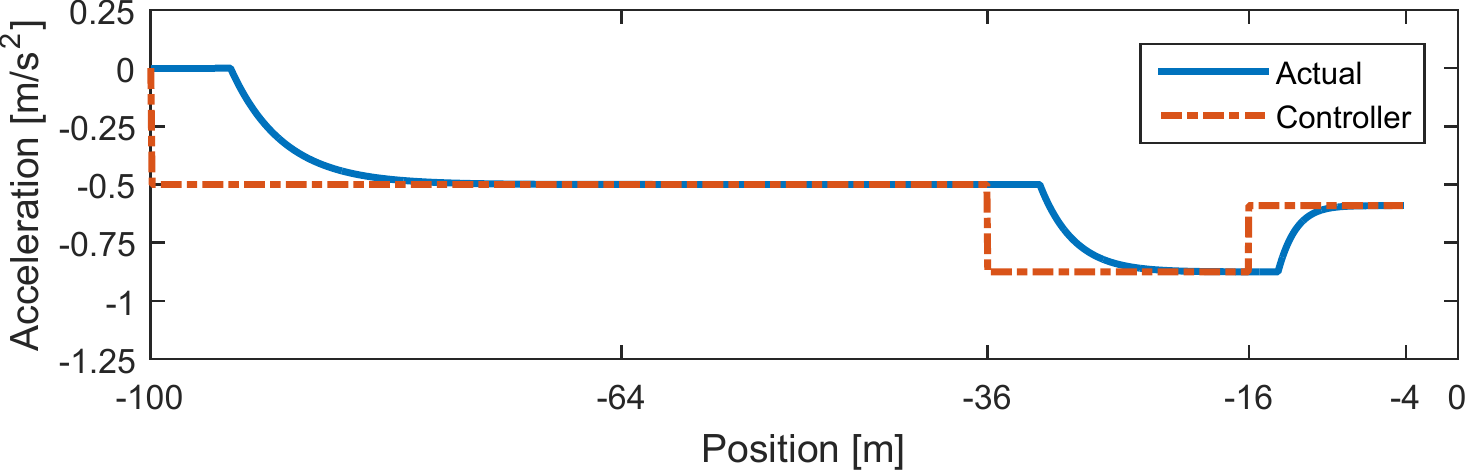}
		\label{fig:same2a}
	}
	\subfigure[Train speed with duplicate attack (maintain accel.).]{
		\includegraphics[width=0.4\linewidth]{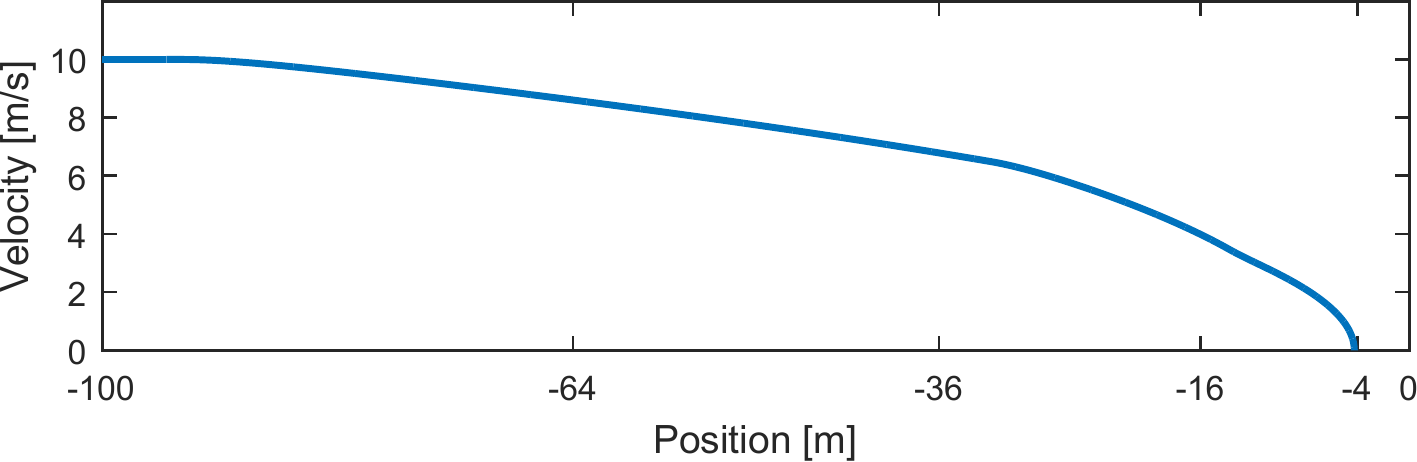}
		\label{fig:same2v}
	}
	\caption{Train acceleration/speed with $B_1$ altered to same as $B_2$.}
	\label{fig:sameattack}
	\vspace{-0.3cm}
\end{figure*}

\subsection{Cloning Attack}
\subsubsection*{{\bf Attack Impact}}
An interesting observation from the HOA braking control model is that the position input from consecutive balises must be different. In Equation~\ref{eq:alphai} if $loc^\prime_i=loc^\prime_{i+1}$ there is a division by zero. It is unclear whether the operational implementations of the HOA algorithm cited in~\cite{Chen2013balise} would be vulnerable, or if additional logic addresses this issue. In our model, we implemented two error handling strategies: (i) the train brakes at full force $\alpha_{max}$ (similar to emergency braking); (ii) the train continues decelerating at its current rate, effectively ignoring the balise.

As was the case in the early stop attack, an adversary wishing to cause a large disruption would tamper with the first few balises, since the train is far from the station and moving quickly. We consider an attack where an attacker tampers with $B_2$ causing $loc^\prime_1=loc^\prime_2=-100$. The results are shown in Figure~\ref{fig:sameattack} for both error handling strategies. In both cases, the train stops before the target point, and outside of the allowable range: Figures~\ref{fig:same1a} and ~\ref{fig:same1v} show a stop at -21.8m in the case with full braking, while Figures~\ref{fig:same2a} and ~\ref{fig:same2v} show a less severe position error of -4.22m for ignoring the repeated input. In both cases, the train would incur additional waiting time at the station, and this additional time would propagate through the system as shown in~\cite{Temple2017}.

\subsubsection*{{\bf Countermeasure Effectiveness}}
Consider the attack shown in Figure~\ref{fig:sameattack}, where the second balise in the stopping sequence is used in a replay attack to impersonate the previous balise. Even after adding balise authentication, this replay attack (copying $B_1$'s data to $B_2$) is still possible. However, by using the train's knowledge about the static (physical) balise deployment and local position estimate in Algorithm~\ref{alg:dtrustworthy} the attack can be detected and corrected.
In the cases where the train's onboard position estimate when it encounters the first balise is $-120m$ or accurate at $100m$, the train obtains a useful data point and therefore knows that the second balise it passes should be $loc_2=-64$ even if the reported value is wrong. As a result, the train stops normally with a small error of $0.15m$. The worst case occurs when the train's initial $p_{est}=-80$. As in the previous attack (Figure~\ref{fig:ccexample}), the train believes it has missed a balise, so it switches to the conservative controller and stops at $0.23m$. 


\section{Conclusion}
\label{sec:conclusion}
In this work we present two countermeasures 
to secure railway spot transmission systems against data integrity and authenticity attacks. The first technique, a low-cost and lightweight authentication method, allows a passing train to detect whether the telegram data provided by a Eurobalise-compliant beacon has been altered. The second countermeasure, a resilient speed controller for train stop control applications, leverages this authentication scheme and adds another layer of security by providing an operationally-viable response strategy when bad data is encountered. Both countermeasures are applicable to legacy systems, in the sense that no additional hardware or sensors are required. 

\section*{Acknowledgments}
 
This work was supported in part by the National Research Foundation (NRF), Prime Minister's Office, Singapore, under its National Cybersecurity R\&D Programme (Award No. NRF2014NCR-NCR001-31) and administered by the National Cybersecurity R\&D Directorate, and supported in part by Singapore's Agency for Science, Technology, and Research (A*STAR) under a research grant for the Human-centered Cyber-physical Systems Programme at the Advanced Digital Sciences Center.

\bibliographystyle{IEEEtran}

\begin{IEEEbiographynophoto}{Hoon Wei Lim} 
is a senior R\&D manager at Cyber Security R\&D, Singtel.
His recent research interests have been centered around data security \& privacy, and security intelligence \& analytics within enterprise environments and cyber-physical systems. 
In the past, he has held research positions at the Institute for Infocomm Research (Singapore), National University of Singapore, Nanyang Technological University, and SAP (France).
Lim received a Ph.D.\ in Information Security from Royal Holloway, University of London.
\end{IEEEbiographynophoto}
\vspace{-10mm}
\begin{IEEEbiographynophoto}{William G. Temple}
received the B.S.\ degree in Mechanical Engineering 
from Tufts University, 
USA, in 2010. He received the M.Eng.\ degree in Mechanical Engineering from Cornell University, 
USA, in 2011. Currently, he is a Project Manager and Senior Research Engineer at the Advanced Digital Sciences Center (ADSC): a research center affiliated with the University of Illinois, located in Singapore. His research interests include cyber security, risk assessment, and resilience of critical infrastructure systems, particularly in the energy and transportation sectors.
\end{IEEEbiographynophoto}
\vspace{-10mm}
\begin{IEEEbiographynophoto}{Bao Anh N. Tran}
received the B.S.\ in Computer Engineering and M.S. in Embedded Systems from Nanyang Technological University in 2005 and 2008, respectively. Currently he is a Software Engineer with Wargaming, Sydney, Australia. 
\end{IEEEbiographynophoto}
\vspace{-10mm}
\begin{IEEEbiographynophoto}{Binbin Chen}
received his B.S.\ from Peking
University, China, in 2003 and his Ph.D.\ from the
National University of Singapore in 2010. He is
currently a Senior Research Scientist at the Advanced
Digital Sciences Center (ADSC), a research center
of the University of Illinois located in Singapore. His
current research interests include wireless networks,
cyber-physical systems, applied algorithms in networking,
and network security. More information
about his research can be found at http://adsc.illinois.edu/people/binbin-chen.
\end{IEEEbiographynophoto}
\vspace{-10mm}
\begin{IEEEbiographynophoto}{Zbigniew Kalbarczyk}
is a research professor at the University of
Illinois, Urbana-Champaign. His research interests include automated design,
implementation, and evaluation of dependable and secure computing systems.
Kalbarczyk received a Ph.D.\ in computer science from the Bulgarian Academy
of Sciences. He is a member of the IEEE and the IEEE Computer Society.
\end{IEEEbiographynophoto}
\vspace{-10mm}
\begin{IEEEbiographynophoto}{Jianying Zhou}
is a professor at the Singapore University of Technology and Design (SUTD).	
His research interests are in applied cryptography, computer and network security, cyber-physical security, mobile and wireless security.
Zhou is a co-founder and/or steering committee co-chair of leading international security conferences, including ACNS, AsiaCCS, and Asiacrypt.
He received a Ph.D.\ in Information Security from Royal Holloway, University of London.
\end{IEEEbiographynophoto}


\end{document}